\documentclass[apl,superscriptaddress,twocolumn,reprint]{revtex4-1}
\usepackage{graphicx}

\begin{document}

\title{Carrier-Envelope Phase Control of a 10 Hz, 25 TW Laser for High-Flux XUV Continuum Generation}

\author{E. Cunningham}
\author{Y. Wu}
\author{Zenghu Chang}
\email{Zenghu.Chang@ucf.edu}
\affiliation{Institute for the Frontier of Attosecond Science and Technology, CREOL and Department of Physics, University of Central Florida, Orlando, FL 32816, USA}

\date{\today}

\begin{abstract}
A novel scheme for stabilizing the carrier-envelope phase (CEP) of low-repetition rate lasers was demonstrated using a 350 mJ, 14 fs Ti:Sapphire laser operating at 10~Hz. The influence of the CEP on the generation of a continuum in the extreme ultraviolet (XUV) was observed.
\end{abstract}

\maketitle

In the field of high-intensity ultrafast laser science, there are certain processes like attosecond photon pulse generation that do not rely directly on the intensity of the femtosecond laser pulse, but rather on the instantaneous amplitude of the underlying electric field oscillations~\cite{Chang2010}. Because of the physical relevance of the sub-cycle character of the laser in such cases, it becomes useful to consider the phase difference between the peak of the pulse's envelope and the closest crest of the pulse's carrier field, which is defined as the carrier-envelope phase (CEP).

In a typical ultrafast oscillator and chirped-pulse amplifier (CPA), there is nothing that naturally constrains the output pulses to have consistent CEP values from shot to shot. In fact, pump energy fluctuations, thermal variations, pointing instabilities, and mechanical vibrations –- especially in the CPA stretcher and compressor -– can all cause the CEP to change in time. Fortunately, researchers over the past decade have devised numerous ways of stabilizing these CEP fluctuations in both the oscillator and the CPA~\cite{Jones2000, Moon2010, Baltuska2003, Chang2006, Li2006OL}. Such stabilization techniques rely on measuring the latest CEP value of a pulse or group of nearby pulses and then providing corrective action to reduce the CEP error compared to the previous measurement.

In the case of a CPA, correction to the CEP can be applied only as often as a new pulse arrives with the latest CEP error information. This means that any CEP jitter occurring at frequencies near to or higher than the laser's repetition rate cannot be measured or compensated. Expressed another way: for a laser with a low repetition rate (e.g. 10~Hz), too many disruptions to the CEP are ``invisible'' to detection because they occur during the relatively long period between consecutive laser pulses, thus resulting in a poor correlation between the CEP of a new pulse and the applied phase corrections determined by the previous pulse. Because of this limitation, CEP-locked CPA systems are all but limited to repetition rates of one kilohertz or higher.

Unfortunately, typical kilohertz CPA systems are themselves limited to pulse energies on the order of millijoules, which restricts the laser's usefulness for studying and controlling strong-field phenomena. If such a laser system is used to generate isolated attosecond pulses, for example, this energy ceiling confines the achievable attosecond pulse energy to tens of nanojoules~\cite{Ferrari2010} –- not enough for performing attosecond pump–-attosecond probe experiments or examining nonlinear XUV dynamics. In order to achieve higher laser intensities and to generate microjoule-level attosecond pulse energies, laser systems must be used that can output joule-level laser pulses. Regrettably, such CPA systems only operate at repetition rates that are too low to be CEP stabilized using existing techniques, making them less ideal for phase-sensitive applications.

In this Letter, we present a ``fast CEP probe'' technique for actively controlling the CEP of a low-repetition rate, high-power CPA laser system for the first time to the best of our knowledge.

\begin{figure}[h!]
  \centering
  \includegraphics{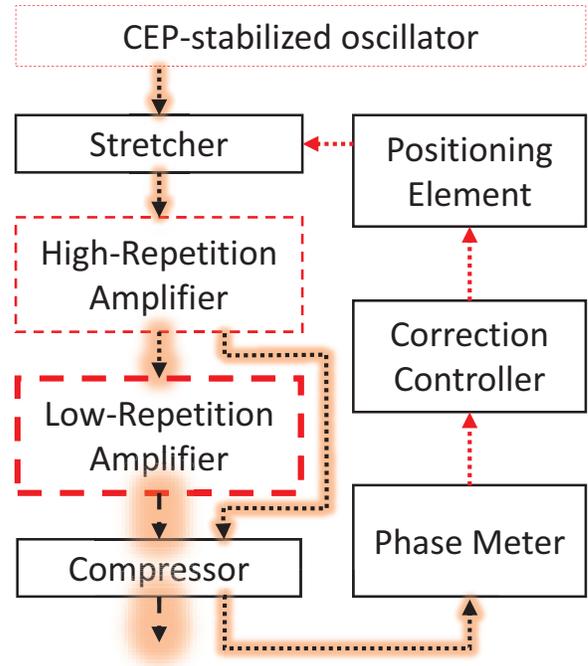}
\caption{The fast CEP probe technique uses a high-repetition sampling beam to measure and control the carrier-envelope phase of a low-repetition amplified beam.\label{fig:block}}
\end{figure}

\begin{figure*}[htb]
  \centering
  \includegraphics{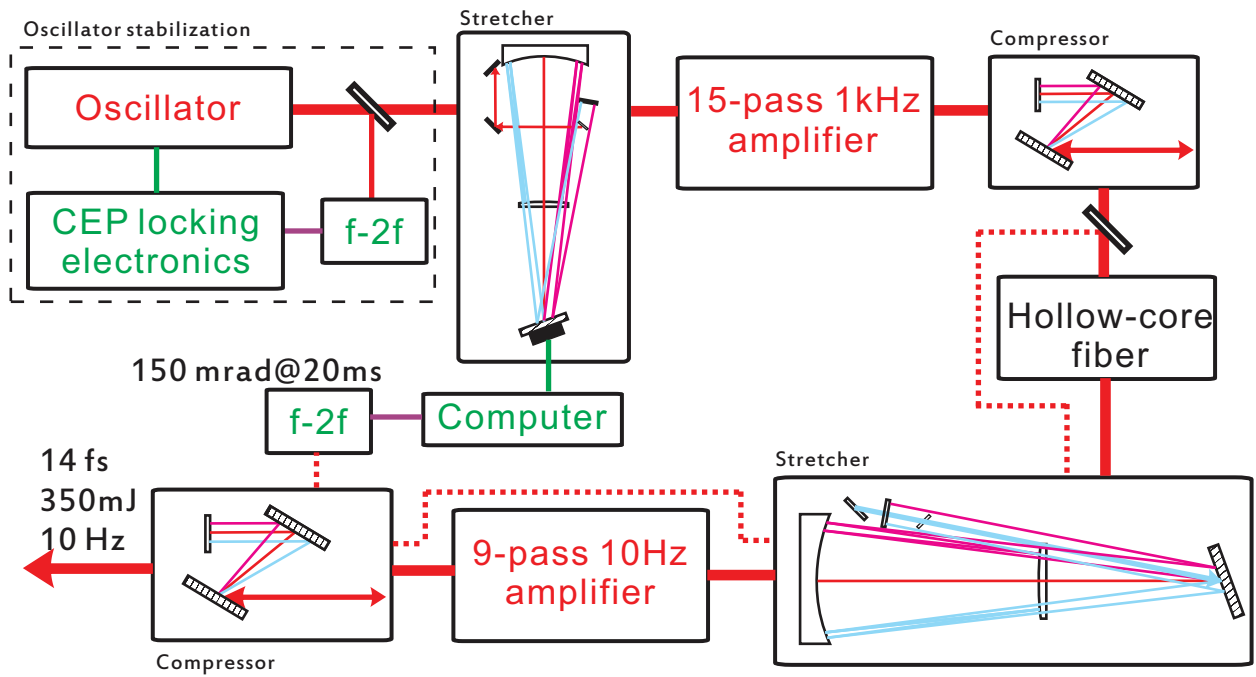}
\caption{The fast CEP probe technique is applied to a 350~mJ, 14~fs (25~TW) Ti:Sapphire laser system.\label{fig:our_laser}}
\end{figure*}

Figure~\ref{fig:block} illustrates the basic concept of this method: the output pulses from a CEP-stabilized oscillator are stretched in time and then amplified at a high repetition rate common to CEP-stabilized systems (e.g. $\geq$1~kHz). While a portion of the beam is further amplified at a low repetition rate atypical to CEP-locked lasers (e.g. $\leq$100~Hz), a small sampling (the ``fast CEP probe'') remains unamplified and maintains the same high repetition rate -- a crucial requirement for providing feedback often enough to lock the CEP. Since the weak, high-repetition pulses and the strong, low-repetition pulses go on to share the same pulse compressor, both beams receive the imprint of the foremost CEP errors arising due to mechanical vibrations and thermal drift in the stretcher and compressor~\cite{Chang2006, Li2006OL}. Because both beams share the same primary CEP error components, the measurement and stabilization of the CEP of the high-repetition probe beam effectively controls the CEP of the low-repetition amplified beam as well.

This technique was first applied to a high-power, Ti:Sapphire-based double CPA system designed for generating high-flux attosecond pulses~\cite{Wu2013}. As represented in Fig.~\ref{fig:our_laser}, the millijoule-level output of the front-end, 1~kHz-repetition rate CPA is spectrally broadened in a hollow-core fiber to provide a broadband seed for the back-end, 10~Hz-repetition rate CPA. To acquire the fast CEP probe beam, a sampling of the 1~kHz beam is taken after the front-end compressor and before the hollow-core fiber. This 1~kHz sampling is directed through both the back-end stretcher {\em and} compressor in order to carry the same primary phase errors as the 10~Hz beam.

Once compressed, the fast CEP probe beam is sent to a home-built f-to-2f interferometer, based on white-light generation in a sapphire plate through filamentation, that is used to measure the CEP~\cite{Kakehata2001}. It is important to note that generating an octave-spanning white-light spectrum strong and stable enough for f-2f interferometry requires input pulses with little shot-to-shot variability and with peak powers greater than the critical power of the sapphire plate. These conditions are fulfilled only after making two special considerations in the back-end CPA:

\begin{enumerate}
\item Because of the changing thermal lensing and residual population inversion inside the Ti:Sapphire crystals in the time between shots of the 10~Hz pump lasers, the 1~kHz fast CEP probe beam must circumvent the 10~Hz multi-pass amplifiers to avoid energy and profile instabilities that would lead to unreliable white-light generation.

\item Because the compressor is tuned to account for the dispersion of the stretcher {\em and the amplifiers} (experienced by the 10~Hz beam only), the 1~kHz fast CEP probe beam must pass through several centimeters of bulk glass directly after the stretcher so that it too will be dispersion-compensated upon entering the f-to-2f interferometer.
\end{enumerate}

\noindent With sufficient white-light generation, the phase retrieved from the 1~kHz interferometer is used to stabilize the CEP of the whole system to 150~mrad RMS (20~ms integration) via feedback control of the grating position~\cite{Li2006OL} in the front-end CPA stretcher.

\begin{figure}[h!]
  \centering
  \includegraphics{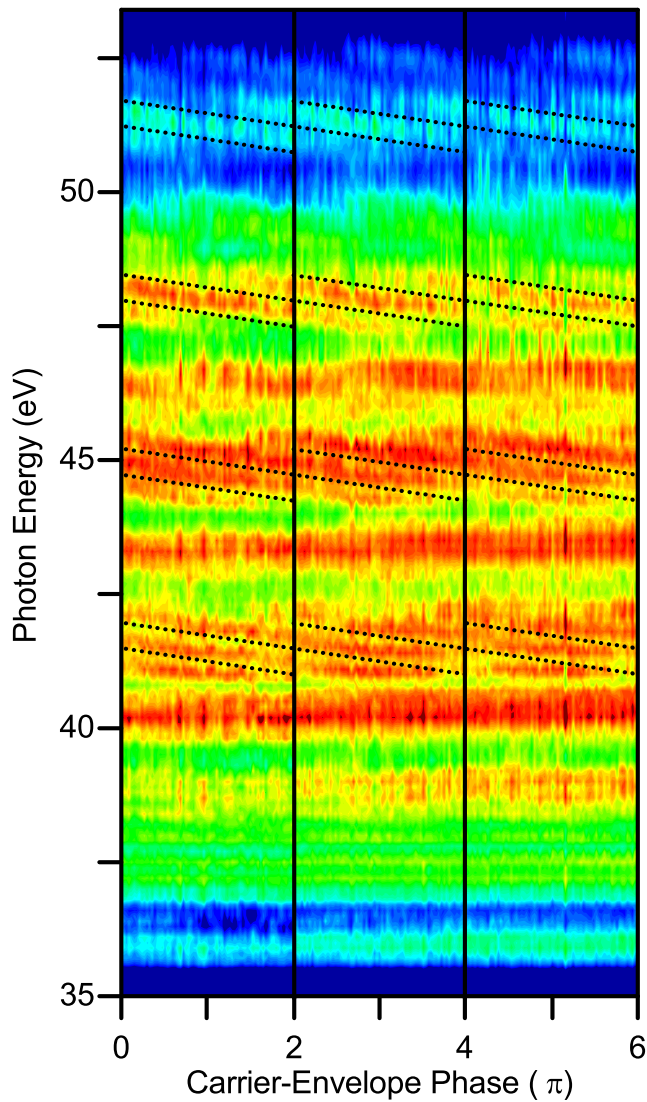}
\caption{Features of an XUV continuum change while scanning the CEP of the 10~Hz, 25~TW driving laser. Dotted lines were added to guide the eyes for easy observation of the 2$\pi$ periodicity of the fine structures. \label{fig:XUV_cont}}
\end{figure}

The efficacy of this locking method for the 10~Hz amplified pulses is gauged by observing the influence of the CEP on the structures of an XUV continuum generated using generalized double optical gating (GDOG)~\cite{Feng2009}. For this experiment, the double CPA system provides 350~mJ, 14~fs pulses at a 10~Hz repetition rate. These pulses propagate through a set of birefringent GDOG optics in order to create an electric field appropriate for isolated attosecond pulse production. After loosely focusing with a concave mirror (f=6.5~m) to a 100 mm-long gas cell containing 4~Torr of argon, the XUV continuum is generated and then separated from the residual near-infrared pulse using two Brewster-angled silicon plates (with a cut-off near 50~eV) and a thin aluminum filter. The isolated XUV spectrum is then measured with a flat-field soft x-ray spectrometer and a charge-coupled device (CCD) detector. As the CEP of the driving laser pulse is scanned linearly from 0 to 6$\pi$~\cite{Li2006OE}, the harmonic spectra varies with a 2$\pi$ periodicity (as shown in Fig.~\ref{fig:XUV_cont}), which is consistent with the periodicity of the two-color GDOG field.

When the amplified 10~Hz beam is sent to an f-to-2f interferometer, the comparatively large shot-to-shot energy and spectrum instabilities of the pulses lead to significant fluctuations in the interferometer's phase retrieval due to changes in the white-light generation process~\cite{Li2009} even when the fast CEP probe beam is locked. Although this has a debilitating effect on the reliability of the interferometer, this directly-measured stability of 1.3~rad RMS nevertheless constitutes an upper bound on the level of CEP control.

In conclusion, we have demonstrated a method for controlling the CEP of a low-repetition, high-power laser. As a result, our 10~Hz, 25~TW CPA system currently represents one of the strongest and lowest-repetition rate CEP-stabilized lasers in the world. This is also one of the first CEP effects seen in an XUV continuum generated using the long gas cell geometry. Our preliminary results pave the way for a new generation of high-field and attosecond science using phase-controlled TW- and PW-level laser sources. This work is funded by the U.S. Army Research Office, the National Science Foundation, and the DARPA PULSE program by a grant from AMRDEC.

\end{document}